\documentclass{ws-procs9x6}
\usepackage{bm}
\usepackage{graphicx}

\newcommand{\gs}{\vert 0\rangle}

\newcommand{\mmc}[1]{\multicolumn{2}{l}{$(#1)\gs$}}

\newcommand{\oc}[1]{$#1\gs$ & }
\newcommand{\tc}[2]{$#1\gs$ & $#2\gs$}
\newcommand{\ooc}[1]{$#1\gs$ }
\newcommand{\ad}[1]{a^\dagger_{#1}}
\newcommand{\aad}[2]{(a^\dagger_{#1})^{#2}}
\newcommand{\pA}[1]{$#1^\prime$}
\newcommand{\pB}[1]{$#1^{\prime\prime}$}
\newcommand{\pC}[1]{$#1^{\prime\prime\prime}$}
\newcommand{\pD}[1]{$#1^{({\rm iv})}$}

\newcommand{\ppA}[1]{$#1^{+\prime}$}
\newcommand{\ppB}[1]{$#1^{+\prime\prime}$}
\newcommand{\ppC}[1]{$#1^{+\prime\prime\prime}$}
\newcommand{\ppD}[1]{$#1^{+({\rm iv})}$}

\begin{document}

\title{Excitations of the static quark-antiquark system
 in several gauge theories\footnote{\uppercase{T}alk presented by 
  \uppercase{C}.~\uppercase{M}orningstar.}}

\author{K.~Jimmy Juge}
\address{School of Mathematics, Trinity College, Dublin 2, Ireland}
\author{Julius Kuti}
\address{Department of Physics, University of California at San Diego,\\
           La Jolla, USA  92093-0319}
\author{Colin Morningstar}
 \address{Department of Physics, Carnegie Mellon University,\\ Pittsburgh,
         PA, USA 15213-3890}

\maketitle

\abstracts{
The spectrum of gluons in the presence of a static
quark-antiquark pair is studied using Monte Carlo simulations on
anisotropic space-time lattices.  For very small quark-antiquark
separations $R$, the level orderings and approximate degeneracies
disagree with the expectations from an effective string
theory.  As the quark-antiquark separation $R$ increases, a dramatic
rearrangement of the energies occurs, and above 2~fm, all of the
levels studied show behavior consistent with an effective string description.
The energy spacings are nearly $\pi/R$, but a tantalizing fine
structure remains.  In addition to 4-dimensional $SU(3)$ gauge theory,
results from 3-dimensional $SU(2)$ and compact $U(1)$ gauge
theories are also presented.}

\section{Introduction}

An important part of understanding confinement in quantum chromodynamics
(QCD) is understanding the low-lying physics of the confining gluon
field.  The spectrum of gluons in the presence of a static
quark-antiquark pair provides valuable clues about the nature of
the low-lying stationary states of the gluon field.  Innumerable lattice
QCD simulations have confirmed that the energy of the ground state
rises linearly with the separation $R$ between the quark and antiquark,
naively suggesting that the gluon field forms a 
string-like confining object connecting the quark and the antiquark.
However, it should be noted that the spherical bag model also predicts
a linearly rising energy for moderate $R$,  and hence, the linearly
rising ground-state energy is {\em not} conclusive evidence of string
formation.  Computations of the gluon action density surrounding a
static quark-antiquark pair in $SU(2)$ gauge theory also hint at flux
tube formation\cite{bali_flux}.  

Adopting the viewpoint that the nature of the confining gluon field
is best revealed in its excitation spectrum, we have embarked on a
series of studies\cite{jkm} employing recent advances in lattice simulation
technology, including anisotropic lattices, improved gauge actions,
and large sets of creation operators, to investigate the onset
of string-like behavior in the gluon field surrounding a quark-antiquark
pair for a wide range of separations $R$ from 0.1 to 3 fm.
Energy gaps given by multiples of $\pi/R$ and a well-defined pattern
of degeneracies and level orderings among the different symmetry
channels form a very distinctive and robust signature of the onset of
the Goldstone modes of the effective QCD string.  Non-universal details
of the underlying string description, such as higher order interactions
and their couplings, are encoded in the fine structure of the spectrum
at large separations.  

In this talk, results from
this series of studies clearly demonstrating the onset of string
formation for large $R$ are presented.  The spectra of three-dimensional
$SU(2)$ and compact $U(1)$ gauge theories are also presented to
address questions about the dependence of results on the gauge
group and the dimensionality of space-time.  First, the classification
of the states is discussed in Sec.~\ref{sec:classify}.  The
expected level orderings at large $R$ from an effective string
description are detailed in Sec.~\ref{sec:string}.  The spectrum
in four-dimensional $SU(3)$ is discussed in Sec.~\ref{sec:su3},
and three-dimensional $SU(2)$ and compact $U(1)$ results are presented
in Sec.~\ref{sec:su2u1}.  A complementary study of the spectrum 
and Casimir energy in three-dimensional $Z(2)$ gauge theory is
presented elsewhere\cite{julius}.

\section{Classification of states}
\label{sec:classify}

The first step in determining the energies of the stationary states
of gluons in the presence of a static quark and antiquark, fixed in
space some distance $R$ apart, is to classify the levels in terms of the
symmetries of the problem.  Such a system has cylindrical symmetry about
the axis $\hat{\bm{R}}$ passing through the quark and the antiquark
(the molecular axis).  The total angular momentum $\vec{\bm{J}_g}$ of the
gluons is not a conserved quantity, but its projection 
$\vec{\bm{J}_g}\!\cdot\hat{\bm{R}}$ onto the molecular axis is and can be
used to label the energy levels of the gluons.  Here, we adopt the standard
notation from the physics of diatomic molecules and denote the magnitude of
the eigenvalue of  $\vec{\bm{J}_g}\!\cdot\hat{\bm{R}}$ by $\Lambda$.
States with $\Lambda=0,1,2,3,4,\dots$ are typically denoted by the capital
Greek letters $\Sigma, \Pi, \Delta, \Phi, \Gamma, \dots$, respectively.  
The energy of the gluons is unaffected by reflections in a plane containing
the molecular axis; since such reflections interchange states of opposite
handedness, given by the sign of the eigenvalue of 
$\vec{\bm{J}_g}\!\cdot\hat{\bm{R}}$, such states must necessarily be
degenerate ($\Lambda$ doubling).  However, this doubling does not apply to
the $\Sigma$ states; $\Sigma$ states which are even (odd) under a reflection
in a plane containing the molecular axis are denoted by a superscript 
$+$ $(-)$.  Another symmetry is the combined operation of charge conjugation
and spatial inversion about the midpoint between the quark and the antiquark.
Here, we denote the eigenvalue of this transformation by $\eta_{CP}$ which
can take values $\pm 1$.  States which are even (odd) under this 
parity--charge-conjugation operation are indicated by subscripts $g$ ($u$).
Thus, the low-lying levels in four space-time dimensions are labeled
$\Sigma_g^+$, $\Sigma_g^-$, $\Sigma_u^+$, $\Sigma_u^-$, $\Pi_g$, $\Pi_u$, 
$\Delta_g$, $\Delta_u$, and so on. 

In three space-time dimensions, there is no longer a rotational symmetry
about the molecular axis since there are only two spatial dimensions.
Instead, the analogous symmetry is a reflection in the molecular axis,
and states are either symmetric $S$ or antisymmetric $A$ under this
transformation. The combined operation of charge conjugation and spatial
inversion about the midpoint between the quark and the antiquark
is still a symmetry in three space-time dimensions.  Once again, states which
are even (odd) under this parity--charge-conjugation operation are indicated
by the subscripts $g$ ($u$).  To summarize, the low-lying states in
three space-time dimensions are labeled by $S_g, A_g, S_u, A_u$.

One last note concerning the classification of states should be made.
In a gauge theory based on the group $SU(2)$, the subscript $g$ and $u$
refers only to spatial inversion about the midpoint between the static
sources, without charge conjugation.  This is due to the fact that in
$SU(2)$, the complex conjugate representation $\overline{\Gamma}$ is
equivalent to the $\Gamma$ representation.

\section{String modes}
\label{sec:string}

The ground-state energy of gluons in the presence of a static quark-antiquark
pair rises linearly with the quark-antiquark separation $R$.  This fact
has led to the general belief that at sufficiently large $R$, the
chromoelectric and chromomagnetic fields become confined to a long
tube-like region of space connecting the quark and the antiquark.  A
treatment of the gluon field in terms of the collective degrees of freedom
associated with the position of the long flux might then be sufficient
for reproducing the long-wavelength physics.  If true, one then hopes that
the oscillating flux can be well described in terms of an effective string
theory.  In such a case, the lowest-lying excitations are expected to be the
Goldstone modes associated with the spontaneously broken transverse
translational symmetry.  These modes are a universal feature of any low-energy
description of the effective QCD string and have energy separations above
the ground state given by multiples of $\pi/R$. For the gluonic excitations
at small $R$, no robust expectations from string theory presently exist.
In this section, the pattern of degeneracies and level orderings of the
expected string modes for large $R$ in both three and four space-time
dimensions are deduced.

\begin{table}[t]
\tbl{Low-lying string levels for fixed ends in four space-time
 dimensions. The $N=1$ level
 is two-fold degenerate, and the $N=2,3,4$ levels are 5,10,15-fold degenerate,
 respectively.  The $+(-)$ signs refer to right (left) circular
 polarizations, and positive integers indicate standing wave normal modes.
\label{table:fixedends}}{
\begin{tabular}{l@{\hspace{10mm}}l@{\hspace{10mm}}l@{\hspace{10mm}}l}\hline
$N=0$:
      & $\Sigma_g^+$   &  \oc{}\\ \hline
$N=1$:
      & $\Pi_u$        &  \tc{\ad{1+}}{\ad{1-}}\\ \hline
$N=2$:
      & \ppA{\Sigma_g} &  \oc{\ad{1+}\ad{1-}}\\
      & $\Pi_g$        &  \tc{\ad{2+}}{\ad{2-}}\\
      & $\Delta_g$     &  \tc{\aad{1+}{2}}{\aad{1-}{2}}\\ \hline
$N=3$:
      & $\Sigma_u^+$   &  \mmc{\ad{1+}\ad{2-}+\ad{1-}\ad{2+}}\\
      & $\Sigma_u^-$   &  \mmc{\ad{1+}\ad{2-}-\ad{1-}\ad{2+}}\\
      & \pA{\Pi_u}     &  \tc{\ad{3+}}{\ad{3-}}\\
      & \pB{\Pi_u}     &  \tc{\aad{1+}{2}\ad{1-}}{\ad{1+}\aad{1-}{2}}\\
      & $\Delta_u$     &  \tc{\ad{1+}\ad{2+}}{\ad{1-}\ad{2-}}\\
      & $\Phi_u$       &  \tc{\aad{1+}{3}}{\aad{1-}{3}}\\ \hline
$N=4$:
      & \ppB{\Sigma_g} &  \oc{\ad{2+}\ad{2-}}\\
      & \ppC{\Sigma_g} &  \oc{\aad{1+}{2}\aad{1-}{2}}\\
      & \ppD{\Sigma_g} &  \mmc{\ad{1+}\ad{3-}+\ad{1-}\ad{3+}}\\
      & $\Sigma_g^-$   &  \mmc{\ad{1+}\ad{3-}-\ad{1-}\ad{3+}}\\
      & \pA{\Pi_g}     &  \tc{\ad{4+}}{\ad{4-}}\\
      & \pB{\Pi_g}     &  \tc{\aad{1+}{2}\ad{2-}}{\aad{1-}{2}\ad{2+}}\\
      & \pC{\Pi_g}     &  \tc{\ad{1+}\ad{1-}\ad{2+}}{\ad{1+}\ad{1-}\ad{2-}}\\
      & \pA{\Delta_g}  &  \tc{\ad{1+}\ad{3+}}{\ad{1-}\ad{3-}}\\
      & \pB{\Delta_g}  &  \tc{\aad{2+}{2}}{\aad{2-}{2}}\\
      & \pC{\Delta_g}  &  \tc{\aad{1+}{3}\ad{1-}}{\ad{1+}\aad{1-}{3}}\\
      & $\Phi_g$       &  \tc{\aad{1+}{2}\ad{2+}}{\aad{1-}{2}\ad{2-}}\\
      & $\Gamma_g$     &  \tc{\aad{1+}{4}}{\aad{1-}{4}}\\ \hline
\end{tabular}}
\end{table}

The excitations of long flux lines are expected to be described by
a local derivative expansion of a massless vector field $\vec{\bm{\xi}}$
with two transverse components in four-dimensional 
space-time\cite{LuescherA,LuescherB}.  Assume that the quark is located
at $(0,0,0)$ and the antiquark is at $(0,0,R)$, then $\vec{\bm{\xi}}(x_3,x_4)$
represents the transverse displacement in the $x_1$ and $x_2$ directions of
the thin flux line from its equilibrium
position.  We further assume that this displacement field is continuous
and single-valued, so that string configurations which double-back on
themselves or overhang the ends are not allowed.
Symmetry arguments then require that the effective QCD string action in
Euclidean space-time should have a leading term given by
\begin{equation}
 S_{\rm eff}^{(0)} = \textstyle\frac{1}{2}c_0\displaystyle
 \int\!\!dx_4\int_0^R\!\!\!dx_3  
 \ \partial_\mu\vec{\bm{\xi}}\cdot
 \partial_\mu\vec{\bm{\xi}},\qquad \vec{\bm{\xi}}(0,x_4)=\vec{\bm{\xi}}(R,x_4)=0,
\end{equation}
where the derivatives are taken with respect to the two worldsheet
coordinates $x_3$ and $x_4$, and $c_0$ has the dimension of a mass squared 
and is proportional to the string tension.  The stationary states are found
by expressing the displacement field $\vec{\bm{\xi}}$ in terms of normal modes. 
For fixed ends, the normal modes are standing waves $\sin(m\pi x_3/R)$. 
These modes have energies $m\omega$
for positive integer $m$ and $\omega=\pi/R$.  For two transverse dimensions,
one defines right $(+)$ and left $(-)$ circularly polarized ladder 
operators
$a_{m\pm}^\dagger$, then the string eigenmodes are
\begin{equation}
\prod_{m=1}^\infty \left(
 \   (a^\dagger_{m+})^{n_{m+}} 
 \  (a^\dagger_{m-})^{n_{m-}} \right) \ \vert 0\rangle,
\end{equation}
where $\vert 0\rangle$ denotes the ground state of the string,
and $n_{m+}$ and $n_{m-}$ are the occupation numbers which take
values $0,1,2,\dots$. 
If $E_0$ denotes the energy of the ground
state, then the eigenvalues $E$ (energy), $\Lambda$, and $\eta_{CP}$
associated with the string eigenstates are given by
\begin{equation}
\begin{array}{rclrcl}
E &=& E_0 + \displaystyle\frac{N\pi}{R}, &\qquad
N&=& \displaystyle\sum_{m=1}^\infty m\ (n_{m+}\!+\!n_{m-}),\\
\eta_{CP} &=& \displaystyle (-1)^N, &\qquad
\Lambda &=& \displaystyle
 \biggl\vert \sum_{m=1}^\infty \left( n_{m+}\!-\!n_{m-}
  \right)\biggr\vert.
\end{array}
\end{equation}
For the $\Sigma$ states, the evenness or oddness under
exchange $(-)\leftrightarrow(+)$ of the
circular polarizations yields a superscript $+$  or $-$, respectively.
Using these properties, the orderings and degeneracies of the Goldstone
string energy levels and their symmetries are
as shown in Table~\ref{table:fixedends}. One sees that the $N\pi/R$ behavior 
and a well-defined pattern of degeneracies
and level orderings among the different channels form a very distinctive
signature of the onset of the Goldstone modes for the effective QCD string.

In three space-time dimensions, there is only one transverse
direction for the string, so the ladder operators are written
$a_m^\dagger$ since there are no right and left circular polarizations.
The orderings and degeneracies of the Goldstone modes are
given in Table~\ref{table:threedim}.

\begin{table}[t]
\tbl{Low-lying string levels for fixed ends in three space-time
 dimensions. The $N=1$ level is nondegenerate, and the 
 $N=2,3,4$ levels are 2,3,5-fold degenerate,
 respectively.  The positive integers indicate the standing wave normal 
 modes.
\label{table:threedim}}{
\begin{tabular}{l@{\hspace{20mm}}l@{\hspace{20mm}}l}\hline
$N=0$:
      & $S_g$        &  \ooc{}\\ \hline
$N=1$:
      & $A_u$        &  \ooc{\ad{1}}\\ \hline
$N=2$:
      & \pA{S_g}     &  \ooc{\aad{1}{2}}\\
      & $A_g$        &  \ooc{\ad{2}}\\ \hline
$N=3$:
      & \pA{A_u}     &  \ooc{\aad{1}{3}}\\
      & \pB{A_u}     &  \ooc{\ad{3}}\\
      & $S_u$        &  \ooc{\ad{1}\ad{2}}\\ \hline
$N=4$:
      & \pB{S_g}     &  \ooc{\aad{2}{2}}\\
      & \pC{S_g}     &  \ooc{\ad{1}\ad{3}}\\
      & \pD{S_g}     &  \ooc{\aad{1}{4}}\\
      & \pA{A_g}     &  \ooc{\ad{4}}\\
      & \pB{A_g}     &  \ooc{\aad{1}{2}\ad{2}}\\ \hline
\end{tabular}}
\end{table}

\section{$SU(3)$ gauge theory in 4 dimensions}
\label{sec:su3}

\begin{figure}[htp]
\centerline{\includegraphics[width=3.8in,bb=50 50 554 770]{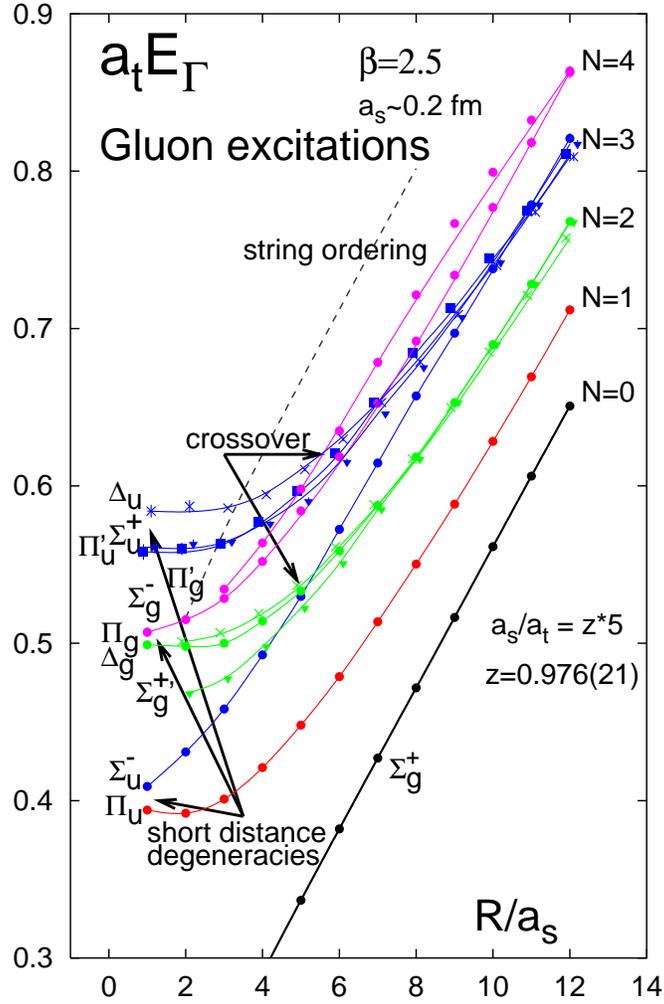}}
\caption{The spectrum of gluonic excitations in the presence of a static
 quark-antiquark pair separated by a distance $R$ in 4-dimensional $SU(3)$
 gauge theory (from Ref.~\protect\cite{jkm}).  Results are from one simulation
 for lattice spacing $a_s\sim 0.2$~fm using an improved action on a
 $(10^2\times 30)\times 60$ anisotropic lattice
 with coupling $\beta=2.5$ and bare aspect ratio $\xi=5$.
 At large distances, {\em all} levels without exception are consistent with
 the expectations from an effective string theory description.
 A dramatic level rearrangement is observed in the crossover
 region between $0.5-2.0$ fm. The dashed line marks a lower bound
 for the onset of mixing effects with glueball states.
\label{fig:su3_spec}}
\end{figure}

\begin{figure}[htp]
\centerline{\includegraphics[width=2.80in,bb=50 50 544 1110]{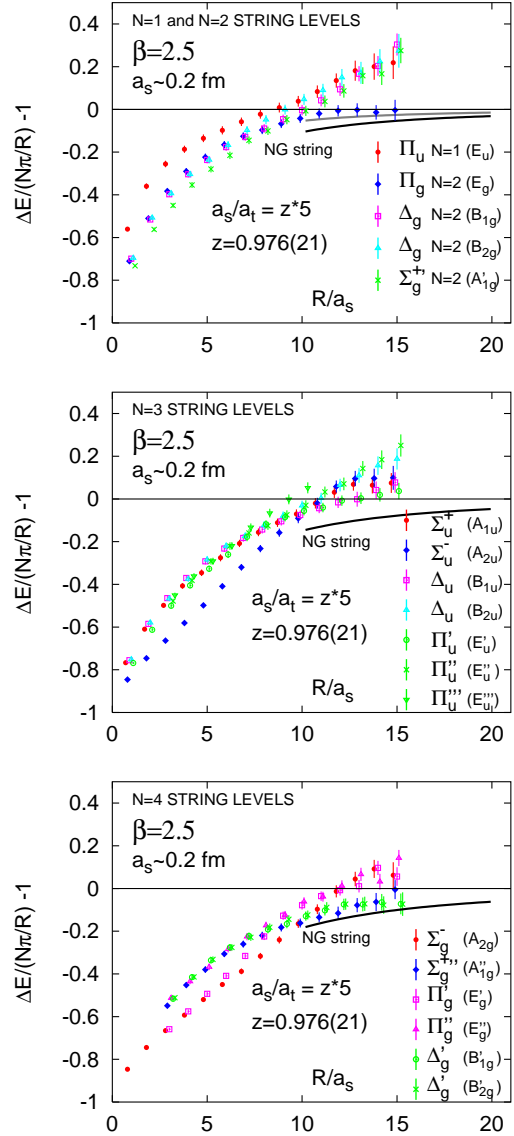}} 
\caption{The energy gaps $\Delta E$ above the ground state
$\Sigma^+_g$ of the stationary states of gluons in the presence of a
static quark-antiquark pair in 4-dimensional $SU(3)$ gauge theory.
The results at lattice spacing $a_s\sim 0.2$~fm are shown against the
quark-antiquark separation $R$ and are compared with the $N\pi/R$ splittings
expected in an effective string theory at large $R$.  The large-$R$
results for a free Nambu-Goto (NG) string are also shown.
\label{fig:su3_gaps}}
\end{figure}

\begin{figure}[ht]
\centerline{\includegraphics[width=2.2in,bb=0 0 730 518]{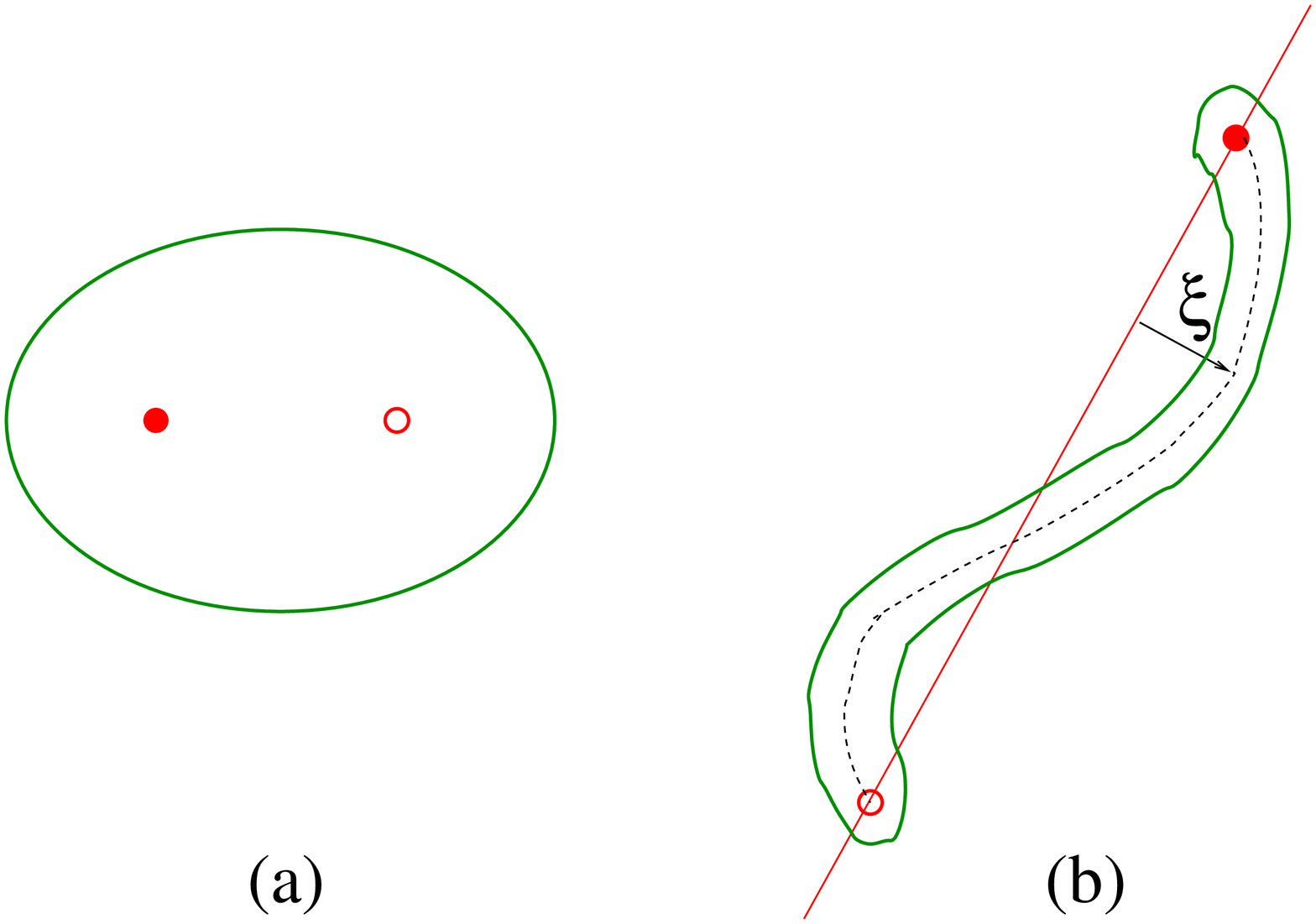}}  
\caption{One possible interpretation of the spectrum in 
 Fig.~\protect\ref{fig:su3_spec}. (a) For small quark-antiquark
 separations, the strong chromoelectric field of the quark-antiquark pair
 repels the physical vacuum (dual Meissner effect) creating a bubble.
 Explanations of the low-lying stationary states must take into account
 both the gluonic modes inside the bubble and oscillations
 of the collective coordinates describing the bubble.
 (b) For large quark-antiquark separations, the bubble stretches into
 a thin tube of flux, and the low-lying states are explained by the
 collective motion of the tube since the internal gluonic excitations are
 much higher lying.
\label{fig:fluxes}}
\end{figure}

The spectrum shown in Fig.~\ref{fig:su3_spec} provides
clear evidence that the gluon field can be well approximated by an
effective string theory for large separations $R$.  Energy gaps $\Delta E$
above the ground state are compared to asymptotic string gaps for 15 excited
states in Fig.~\ref{fig:su3_gaps}.  The quantity $\Delta E/(N\pi/R)-1$
is plotted to show percentage deviations from the asymptotic string
levels for string quantum number $N=1,2,3,4$.  For small $R<2$~fm,
the energy gaps lie far below the null lines and are strongly split for
fixed $N$.   In other words, string formation does not appear
to set in until the quark and the antiquark are separated by about 2 fm.  
For small separations, the level orderings and
degeneracies are not consistent with the expectations from an effective string
description.  More importantly, the gaps differ appreciably from $N\pi/R$
with $N=1,2,3,\dots$, as clearly shown in Fig.~\ref{fig:su3_gaps}.
Such deviations, as large as $50\%$ or more, cannot be considered mere
corrections, making the applicability of an effective string description
problematical.  Between 0.5 to 2 fm, a dramatic level rearrangement occurs.
For separations above 2 fm, the levels agree {\em without exception} with
the ordering and degeneracies expected from an effective string theory.
The gaps agree well with $N\pi/R$, but a fine structure remains.  This first
glimpse of such a fine structure offers the exciting possibility of
deducing details of the effective QCD string action in future
higher precision simulations.

It is reasonable to expect that the first few terms in the effective
string action might predominantly arise from the geometric properties
of the flux tube.  The Nambu-Goto (NG) action is one of the simplest
geometrical string models.  The spectrum of the Nambu-Goto string
with fixed ends in $d$ dimensions has been calculated\cite{arvis},
with the result
\begin{equation}
 E_N=\sigma R\left(1-\frac{(d-2)\pi}{12\sigma R^2}+\frac{2\pi N}{\sigma
 R^2}\right)^{1/2}.
\end{equation}
For small $R$, this model has a quantization problem\cite{arvis}
unless $d=26$, but the problem disappears as $R$ becomes large.
The energy gaps expected for a Nambu-Goto string at large $R$ are
shown in Fig.~\ref{fig:su3_gaps}.  Deviations of the simulation results
from the Nambu-Goto gaps suggests that physical properties,
such as rigidity, may be relevant for the
effective string action.

Fig.~\ref{fig:fluxes} illustrates one possible interpretation of the
results shown in Fig.~\ref{fig:su3_spec}.  At small quark-antiquark
separations, the strong chromoelectric field of the quark-antiquark pair
repels the physical vacuum in a dual Meissner effect, creating a bubble
surrounding the pair.   Descriptions of the low-lying stationary states
must take into account both the gluonic modes inside the bubble and
the motion of the collective coordinates describing the bubble.
For large quark-antiquark
separations, the bubble stretches into a thin tube of flux, and the low-lying
states could then be explained by the collective motion of the tube since
the internal gluonic excitations, being typically of order 1 GeV, are
now much higher lying.  We caution the reader that the above interpretation
is simply speculation based on observations to date.  Although the simulation
results rule out the usefulness at small $R$ of an effective string action
constructed as a $1/R$ expansion, they do not actually rule out the unlikely
possibility of a string description based on some other expansion parameter.

\section{$SU(2)$ and compact $U(1)$ gauge theories in 3 dimensions}
\label{sec:su2u1}

\begin{figure}[htp]
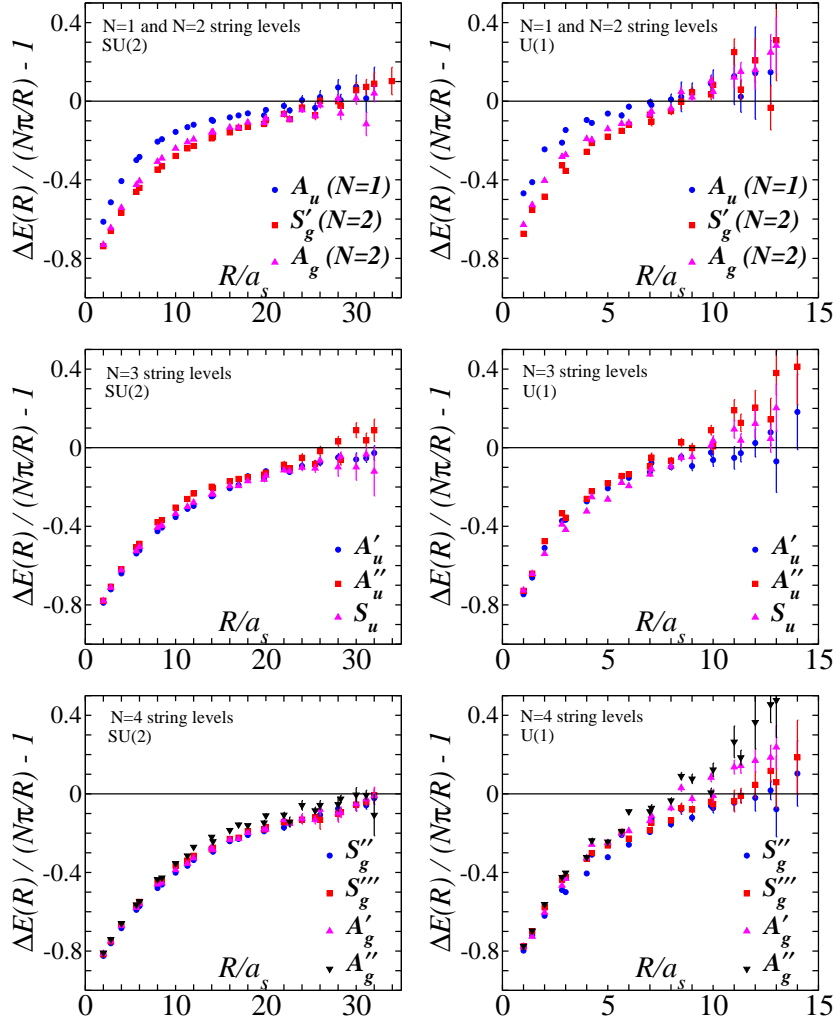

\centerline{\includegraphics[width=2.0in,bb=29 62 584 523]{su2_gapsA}
\hspace*{1em}\includegraphics[width=2.0in,bb=29 62 584 523]{u1_gapsA}}
\vspace*{1em}
\centerline{\includegraphics[width=2.0in,bb=29 62 584 523]{su2_gapsB}
\hspace*{1em}\includegraphics[width=2.0in,bb=29 62 584 523]{u1_gapsB}}
\vspace*{1em}
\centerline{\includegraphics[width=2.0in,bb=29 62 584 523]{su2_gapsC}
\hspace*{1em}\includegraphics[width=2.0in,bb=29 62 584 523]{u1_gapsC}}  
\caption{The energy gaps $\Delta E$ above the ground state
$S_g$ of the stationary states of the gauge field in the presence of a
static source pair in 3-dimensional $SU(2)$ and compact $U(1)$ gauge theories.
These gaps are shown against the separation $R$ of the static sources
and are compared with the $N\pi/R$ splittings expected in an effective
string theory at large $R$.  The results in $SU(2)$ were obtained
on an $84^3$ lattice using an anistropic improved lattice action
with coupling $\beta=5.6$ and bare aspect ratio $\xi=2$, so that 
$a_s\sim 0.1$~fm.  The compact $U(1)$
results were obtained on a $28^2\times 224$ lattice using an anisotropic
improved lattice action for $\beta=0.5$ and $\xi=8$.
\label{fig:su2_gaps}}
\end{figure}

The spectra of three-dimensional $SU(2)$ and compact $U(1)$ gauge theories
were also studied to address questions about the dependence of the results on
the gauge group and the dimensionality of space-time.  Due to the
reduced dimensionality, higher statistical precision was possible in
these calculations.  These simulations also served to check various
systematic errors.

The excitation gaps $\Delta E$ above the $S_g$ ground state of ten
levels were computed and are compared with $N\pi/R$ in
Fig.~\ref{fig:su2_gaps}.  Again, the large-$R$ results are consistent with
the expectations from an effective string description without exception.
A fine structure is also observed, but it is less pronounced than
that in four-dimensional $SU(3)$.  Unlike in four-dimensional $SU(3)$,
no dramatic level rearrangements occur between small and large separation,
but deviations from $N\pi/R$ are significant for small $R$.  There is
remarkable agreement between the $SU(2)$ and compact $U(1)$ results.
A detailed examination of these results is still work in progress.

We have also pursued the spectrum in three-dimensional $Z(2)$ lattice
gauge theory.  These results are reported elsewhere\cite{julius}.
Extremely precise determinations are possible in $Z(2)$ by exploiting
a duality transformation into an Ising model.  In the critical region,
the resulting Ising model admits a description in terms of a $\phi^4$ real
scalar field theory, allowing the possibility of understanding the
underlying microscropic origins of confinement in a rigorous
field-theoretical setting.  The details of this work in progress
are presented elsewhere\cite{julius}.

\section{Conclusion}

In this talk, Monte Carlo computations of the energies of sixteen
stationary states of the gluon field in the presence of a static
quark-antiquark pair separated by a distance $R$ were presented
for a wide range of $R$ from 0.1 to 3~fm.
Striking confirmation of string-like flux formation
of the gluon field surrounding a quark-antiquark pair separated by
distances larger than 2~fm was presented.  A tantalizing fine structure
was revealed, suggesting the possibility of identifying the effective
QCD string action in future higher precision simulations.  A dramatic
level rearrangement between small and large quark-antiquark separations
was observed in a crossover region around 2~fm.  The observed pattern
of energy levels at small $R$ strongly challenges an effective string
description.

Eleven levels in three-dimensional $SU(2)$ and compact $U(1)$ lattice
gauge theory were also studied.  String formation was once again
confirmed at large separations, with a fine structure less pronounced
than in four-dimensional $SU(3)$.  No dramatic level rearrangement
was found between large and small separations.  These studies
are ongoing, and we are also vigorously pursuing the spectrum
and other observables in three-dimensional $Z(2)$ gauge theory with
the goal of determining the effective string action.
Future work also includes calculating the 
three-dimensional $SU(3)$ spectrum, torelon (flux loops winding
around the lattice) spectra, and studying the spatial structures
of these gluonic excitations.
This work was supported by the 
U.S.\ National Science Foundation under award PHY-0099450,
the U.S.\ DOE, Grant No. DE-FG03-97ER40546, and the European
Community's Human Potential Programme under contract HPRN-CT-2000-00145,
Hadrons/Lattice QCD.


\begin{thebibliography}{9}

\bibitem{bali_flux}
G.S.~Bali, K.~Schilling, C.~Schlichter, Phys.\ Rev.\ D {\bf 51}, 5165 (1995).

\bibitem{jkm}
K.J.~Juge, J.~Kuti, and C.~Morningstar, Phys.\ Rev.\ Lett.\ {\bf 90}, 161601
  (2003).

\bibitem{julius}
 K.J.~Juge, J.~Kuti, and C.~Morningstar, these proceedings.

\bibitem{LuescherA}
 M.~L\"uscher, K.~Symanzik, P.~Weisz,
 Nucl.~Phys.\ {\bf B173}, 365 (1980).
   
\bibitem{LuescherB}
 M.~L\"uscher,
 Nucl.~Phys.\ {\bf B180}, 317 (1981). 

\bibitem{arvis} 
 J.~F.~Arvis, Phys.\ Lett.\ {\bf 127B}, 106 (1983).

\end{thebibliography}
\end{document}